\documentclass{aastex631}

\usepackage{graphicx}
\usepackage{gensymb}
\usepackage{csvsimple}
\shorttitle{Flux Calibration and Non-linear Reprojection}
\shortauthors{Kharchilava et al.}

\begin{document}

\title{CARRSSPipeline: Flux Calibration and Non-linear Reprojection for SALT-RSS Multi-Object Spectroscopy over 3500-9500 {\AA}}

\author[0000-0002-0927-7554]{George V. Kharchilava}
\affiliation{Department of Physics \& Astronomy \\
Rutgers, The State University of New Jersey \\
Piscataway, NJ 08854, USA}

\author{Eric Gawiser}
\affiliation{Department of Physics \& Astronomy \\
Rutgers, The State University of New Jersey \\
Piscataway, NJ 08854, USA}

\author{Matt Hilton}
\affiliation{Wits Centre for Astrophysics \\
University of the Witwatersrand, Johannesburg \\
Braamfontein, Johannesburg, 2000, South Africa}

\author{Elisabeth Turner}
\affiliation{Department of Physics \& Astronomy \\
Tufts University \\
Medford, MA 02155, USA}

\author{Nicole Firestone}
\affiliation{Department of Physics \& Astronomy \\
Rutgers, The State University of New Jersey \\
Piscataway, NJ 08854, USA}

\author{Kyoung-Soo Lee}
\affiliation{Department of Physics \& Astronomy \\
Purdue University \\
West Lafayette, IN 47907, USA}



\begin{abstract}

The Robert Stobie Spectrograph (RSS) on the Southern African Large Telescope (SALT) offers multi-object spectroscopy over an 8\arcmin{} field-of-view at resolutions up to 
R $\sim$ 3000.  
Reduction is typically conducted using \texttt{RSSMOSPipeline}, which performs basic data calibrations, sky subtraction, and wavelength calibration.  
However, flux calibration of SALT-RSS using spectrophotometric standard star observations 
is difficult due to variable primary mirror illumination. We describe a novel approach where stars with Sloan Digital Sky Survey spectra are included as alignment stars on RSS slitmasks and then used 
to perform flux calibration of the resulting data.   
RSS offers multiple settings that can be pieced together to cover the entire optical range, utilizing grating angle dithers to fill chip gaps. We introduce a non-linear reprojection routine that defines an exponential wavelength array spanning 3500-9500~{\AA} with gradually decreasing resolution and then reprojects 
several 
individual settings 
into a single 2D spectrum
for each object. 
Our flux calibration and non-linear reprojection routines 
are released as part of the Calibration And Reprojection for RSS Pipeline (\texttt{CARRSSPipeline}; https://github.com/GeorgeTheGeorgian/CARRSSPipeline.git),
that enables the extraction of full-optical-coverage, flux-calibrated, medium-resolution one-dimensional spectra.     

\end{abstract}
\submitjournal{Publications of the Astronomical Society of the Pacific}


\section{Introduction \label{sec:intro}}
Nebular emission science involves the identification and investigation of emission lines and their fluxes, line ratios, and observed wavelengths. These can be used to determine the physical properties of the galaxies they originate from, including electron densities, ionization parameters, velocity dispersion, dust reddening, metallicities, and star formation rates. In order to accurately determine these characteristics, proper reduction must be carried out on observed spectroscopic data to remove or correct for both instrument systematics and external factors. Wavelength and flux calibration in particular are needed, with the latter posing challenges due to instrument limitations (see Section \ref{subsec:flux}). Analyses of velocity dispersion and metallicity can be carried out prior to flux calibration and dust correction since these only require measurements of line widths or flux ratios of lines that are close to each other. However, accurate fluxes are required for flux ratios of well-separated lines, as well as determining other characteristics such as star formation rates. The Multi-Object Spectroscopy (MOS) mode on the Robert Stobie Spectrograph (RSS) on the Southern African Large Telescope (SALT) offers full optical wavelength coverage across several gratings, as well as resolutions of up to R $\sim$ 3000 over an 8\arcmin{} field-of-view. This allows us to resolve the roughly 2.7~\AA{} separated [O II] doublets, and measure each central wavelength and line flux of [O II]$\lambda$3726 and [O II]$\lambda$3729. 

There are several spectroscopic reduction pipelines made in Python\footnote{https://www.python.org/} that are available for astronomers to use, such as \texttt{PyReduce}\footnote{https://pyreduce-astro.readthedocs.io/en/latest/index.html} \citep{2021A&A...646A..32P}, \texttt{PyDIS}\footnote{https://github.com/StellarCartography/pydis}, \texttt{specreduce}\footnote{https://github.com/astropy/specreduce}, and \texttt{Pypelt}\footnote{https://pypeit.readthedocs.io/en/release/} \citep{pypeit:joss_arXiv,pypeit:zenodo}. However, some of these packages are only available for longslit spectroscopy while others only perform up to wavelength calibration. Even the versatile \texttt{Pypelt}, which does perform flux calibration, does not include an RSS instrument package. That is why we seek to develop a pipeline using Python routines to carry out full data calibration, called the Calibration And Reprojection for the Robert Stobie Spectrograph Pipeline (\texttt{CARRSSPipeline}\footnote{https://github.com/GeorgeTheGeorgian/CARRSSPipeline.git}). These calibrations pose challenges due to the nature of both the instrumentation being used and the condition of the observations as discussed in Section \ref{sec:obsIns}, and we pick relatively high flux targets for our masks as priority objects together with dim filler targets, and alignment stars. 
Section \ref{sec:RSSMOS} introduces the first steps in the reduction process, which are done by a preceding Python pipeline. We describe data handling and reduction processes of the \texttt{CARRSSPipeline} in Section \ref{sec:CARRSS}, including the results of our novel approach to flux calibration and its statistics. Future directions and further developments are mentioned in Section \ref{sec:conc}.

\section{Observations \label{sec:obsIns}}
Once SALT-RSS MOS data is collected and downloaded, it is subject to two pipelines to carry out all reduction steps: \texttt{RSSMOSPipeline}\footnote{https://\texttt{RSSMOSPipeline}.readthedocs.io} (see Section \ref{sec:RSSMOS}) which performs flat-field correction, cosmic ray rejection, sky subtraction, and wavelength calibration, and our \texttt{CARRSSPipeline} (see Section \ref{sec:CARRSS}) which performs flux calibration, sky subtraction correction, wavelength reprojection, continuum comparison, and line inspection. These pipelines are made in Python and are designed specifically to handle SALT-RSS MOS data. The second pipeline directly follows the first and outputs both 2D and 1D flux-calibrated, fully reprojected spectra.

\begin{figure}
    \centering
    \plotone{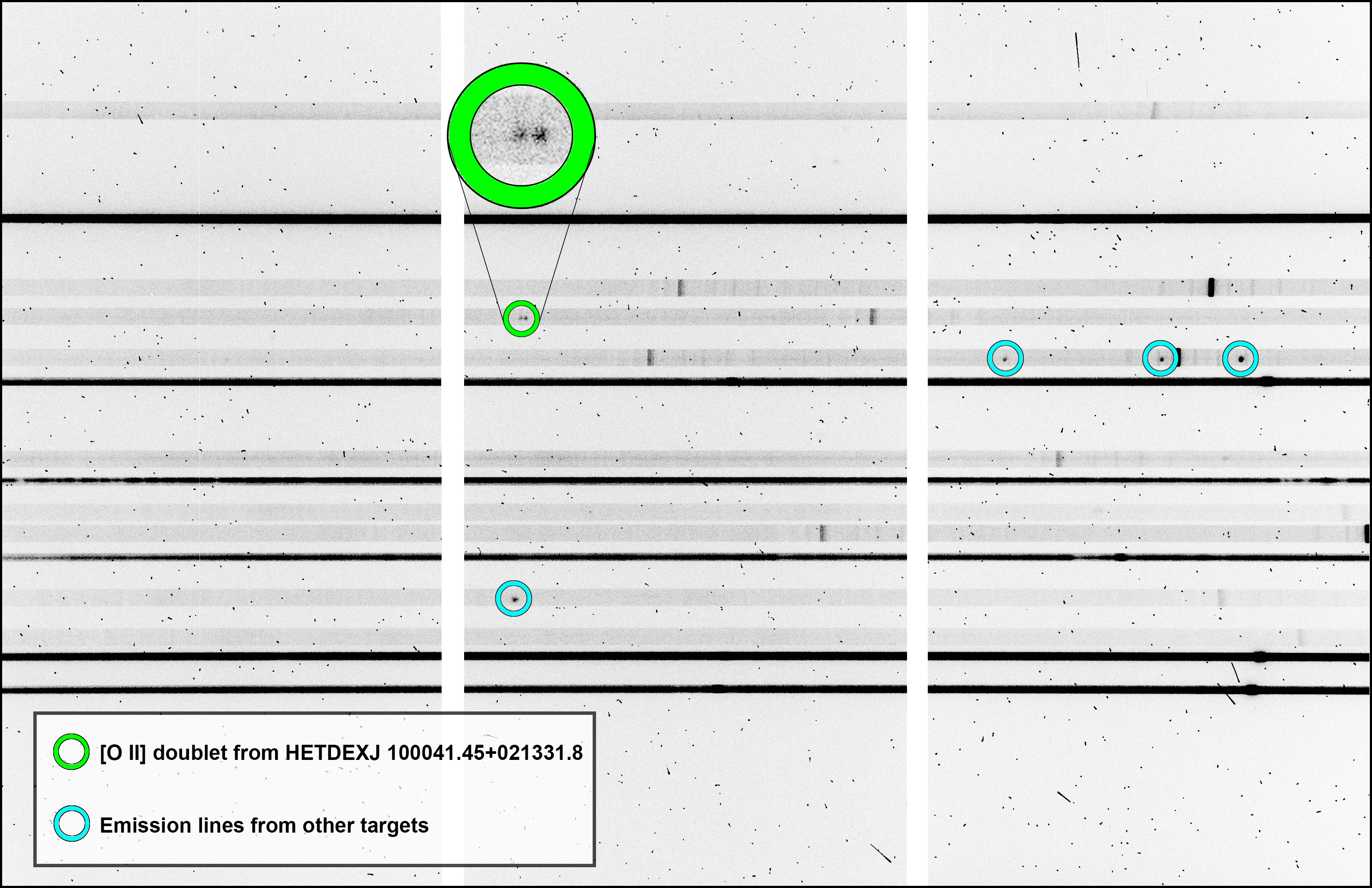}
    \caption{Science exposures of COSMOS-Mask-B taken with grating PG2300 at the lower central wavelength with some emission lines circled. Each horizontal slit is a separate target, and the fourth slit from the top with its [O II] doublet circled in green is J100041.45+021331.8, a HETDEX [O II]-emitter. Lines from other targets, such as [O III] and H$\beta$, are circled in light blue. The six darkest spectra correspond to alignment star boxes.}
    \label{fig:three_subfigures}
\end{figure}

Our mask labeled COSMOS-mask-B was created using the \texttt{PySALT}\footnote{https://github.com/saltastro/pysalt.git} \citep{10.1117/12.857000} user package for the SALT telescope, and observed during the 2021-2 semester as program 2021-2-SCI-026 (Principal Investigator Elisabeth Turner). The mask included emission-line targets at redshift z $<$ 0.4 from the Hobby-Eberly Telescope Dark Energy Experiment survey \citep[HETDEX:][]{1998SPIE.3352...34R,2021ApJ...923..217G,2021AJ....162..298H}. Though [O II] is most commonly seen in our setting using the PG3000 grating, Figure \ref{fig:three_subfigures} shows a PG2300 setting for COSMOS-mask-B with an [O II] doublet, as well as an [O III] doublet and H$\beta$ emission. We also identify targets with H$\alpha$ and [N II] lines in PG0900 for the same mask. In this paper, we follow one of these targets, called HETDEX J100041.45+021331.8, throughout the reduction process, with the purpose of demonstrating accurate calibrations as well as maintaining the resolution of the [O II] doublet. The gratings, wavelength ranges, and central resolution values chosen for our program are PG3000 ($\lambda\lambda$ = 3409-4488~{\AA}, R $\sim$ 1936, 2188), PG2300 ($\lambda\lambda$ = 4197-5506~{\AA}, R $\sim$ 1956, 2167\footnote{Since PG2300 Dither+ was not observed in our COSMOS-mask-B, this value was measured using observations from another mask.}), \& PG0900 ($\lambda\lambda$ = 5191-9530~{\AA}, R $\sim$ 902, 1096), for 2\arcsec{} slit widths with a corresponding resolution element of 7.87 pixels. With two central wavelengths per grating, our program has a total of six spectroscopic settings that observe each mask. We will denote each dither as Dither+ for the higher central wavelength and Dither$-$ for the lower central wavelength. The astronomer's log recorded clear weather conditions during all observations for this mask. Seeing was 0.98\arcsec{} for the PG3000 Dither$-$,  1.5\arcsec{} for PG3000 Dither+, 1.1\arcsec{} for PG2300 Dither$-$, 1.2\arcsec{} for PG0900 Dither$-$, and 1.9\arcsec{} - 2.4\arcsec{} for PG0900 Dither+. For COSMOS-mask-B, the PG2300 Dither+ was not observed due to poor observing conditions. As a result, some chip gaps in this example have not been filled. However, the successful reduction of HETDEX J100041.45+021331.8 outlines the \texttt{CARRSSPipeline}'s capabilities of reprojecting data even when settings are missing (see Section \ref{sec:CARRSS}).

\clearpage
\section{RSSMOSPipeline \label{sec:RSSMOS}}
We employ \texttt{RSSMOSPipeline} \citep{Hilton_2018} to carry out the preliminary reduction processes, which includes flat-field correction, cosmic ray rejection, sky subtraction, and wavelength calibration. Two of the main components of the package are the arc model generator and the reducer, which runs the scripts. \texttt{RSSMOSPipeline} performs these reductions on each spectroscopic setting separately, and outputs the 2D spectra as a Flexible Image Transport System (FITS) file. 

While wavelength calibration is a standard procedure in spectroscopic reduction, our specific program requires reasonably accurate ($\sim$1~\AA) wavelength agreement across multiple settings. The precision of this calibration influences subsequent steps, such as wavelength reprojection used for spectroscopic combination as detailed in Section \ref{subsec:speccombine}.
The output diagnostic plots from \texttt{RSSMOSPipeline} help the user visually inspect the wavelength solution and the features used to make them. The more features used from our model, the more accurate the wavelength solution. This can, however, become an issue when the arc image lacks noticeable features at low dispersion.
Despite this, the wavelength solution shows that the centroid of the [O II] doublet from the PG2300 setting appears at $\lambda$ = 4938.9~\AA, which matches the HETDEX reported wavelength of $\lambda$ = 4938.38~\AA, to within an Angstrom. We can further show the reliability of our wavelength solutions across other gratings by using the redshift calculated from the doublet and seeing if it is consistent with other lines. The results are consistent, with the redshift coming out to z = 0.325 for each line. Results for overall wavelength calibration accuracy are reported in Section \ref{subsec: line}.

\section{CARRSSPipeline\label{sec:CARRSS}}
We develop \texttt{CARRSSPipeline}, a Python routine that compliments \texttt{RSSMOSPipeline}, to perform flux calibration, sky subtraction correction, wavelength reprojection, continuum comparison, and line inspection. This routine is designed to take several spectroscopic settings and combine them post calibration to cover a wavelength range of 3500~\AA{} to 9500~\AA{} and can be expanded or scaled down to accommodate as many settings as desired. The data must first be manually organized to place each mask in a separate directory. Each mask directory will contain sub-directories for each grating and within those, sub-directories for each grating angle. Once properly organized, reduction routines are carried out one mask at a time. 

\subsection{Flux Calibration\label{subsec:flux}}
Flux calibration on SALT-RSS is considered highly challenging due to time variable primary mirror illumination resulting from the tracker moving as the sky rotates \citep{RomeroColmenero2023}. However, we develop a novel approach using Sloan Digital Sky Survey (SDSS) stars as standards for flux calibration. In addition to the mask directories, an SDSS star sub-directory must also be manually made that contains all SDSS spectra of SDSS alignment stars for a given mask. The \texttt{CARRSSPipeline} uses these directories as inputs for its \texttt{flux\_calibration} function.

\begin{figure}[ht]
    \plottwo{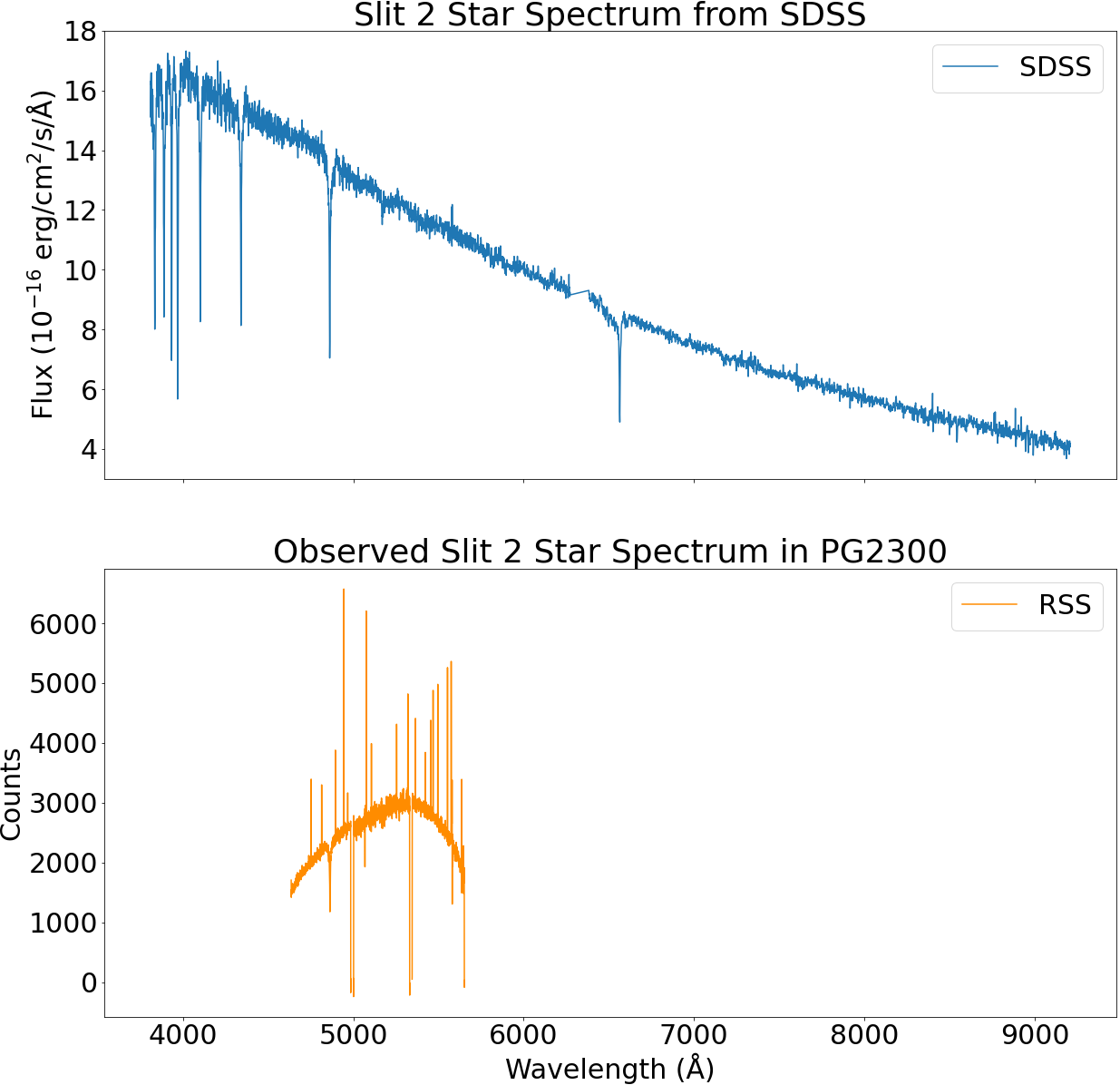}{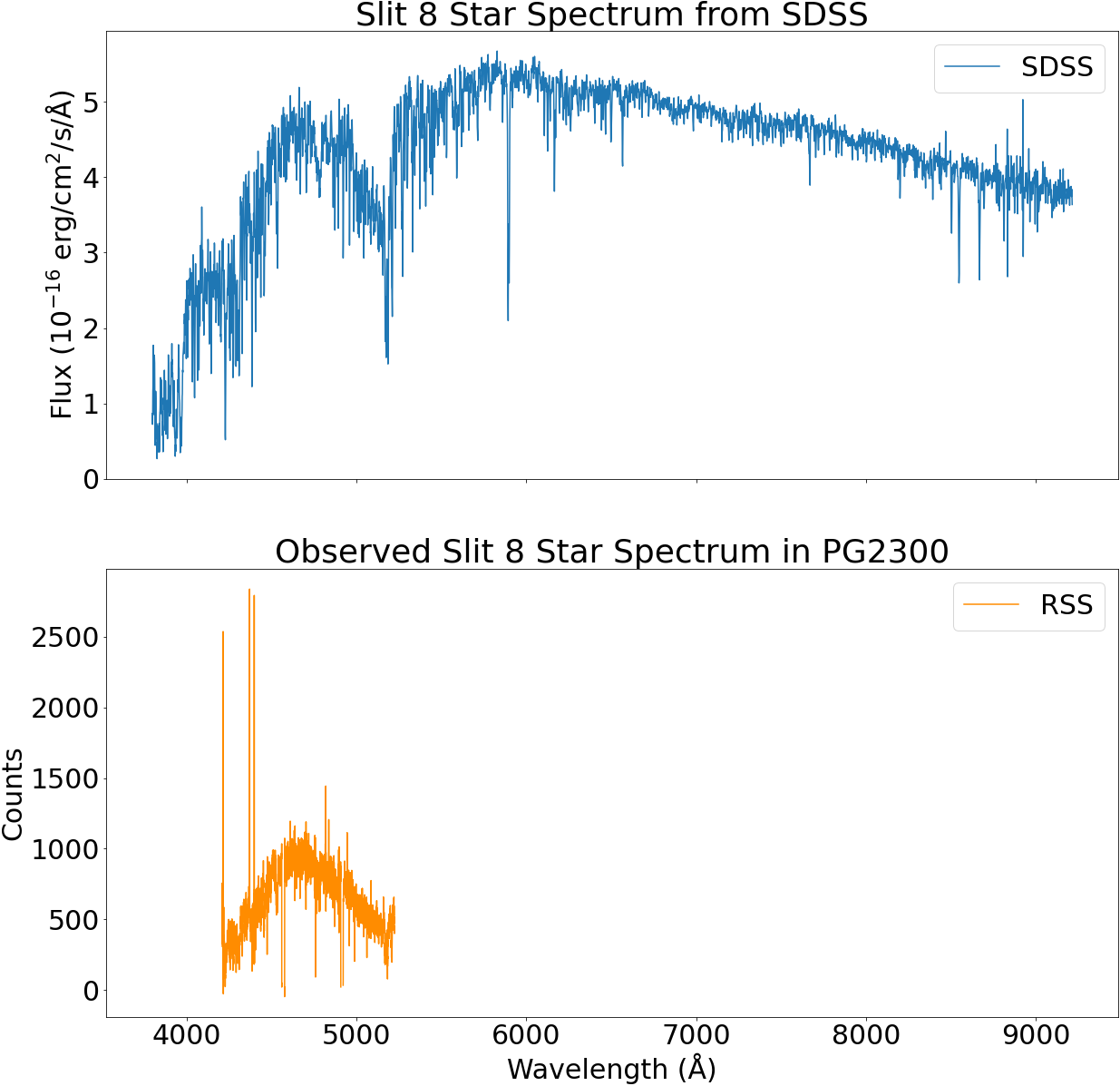}
    \caption{Plots for spectra from SDSS (upper/blue) and observed by RSS (lower/orange) of two alignment stars in COSMOS-mask-B. This particular mask has only two reliable alignment stars with SDSS spectra, but the routine can accommodate an arbitrary number of stars.}
  \label{fig:obsvssdss}
\end{figure}




\begin{figure}[ht]
    \plottwo{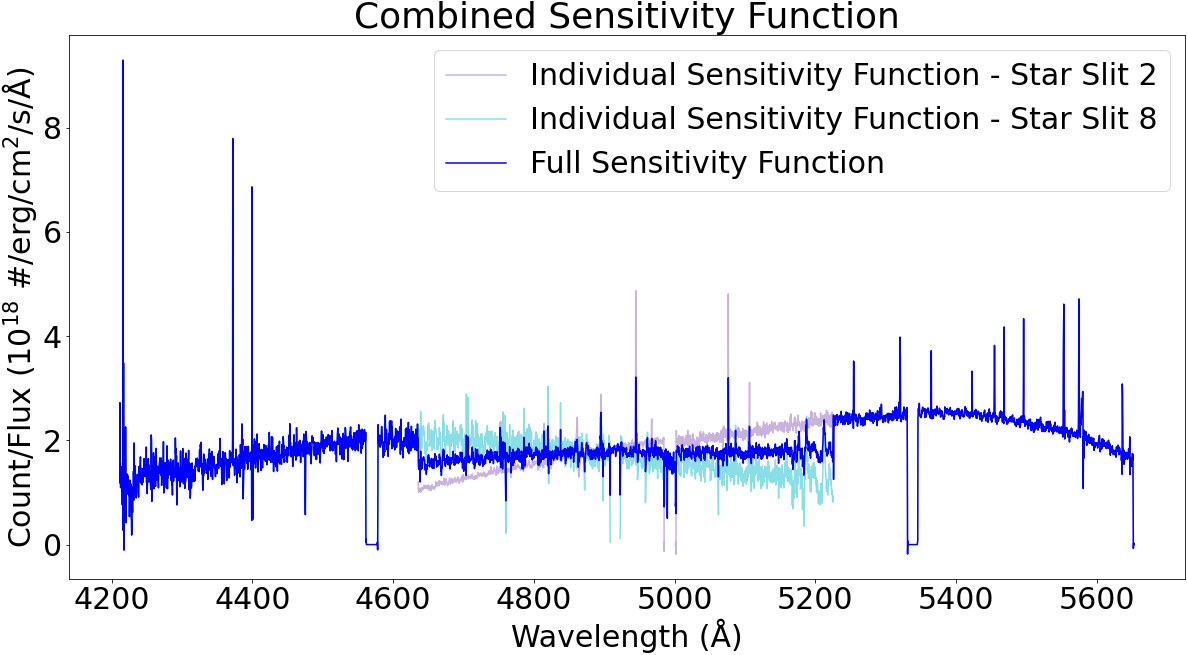}{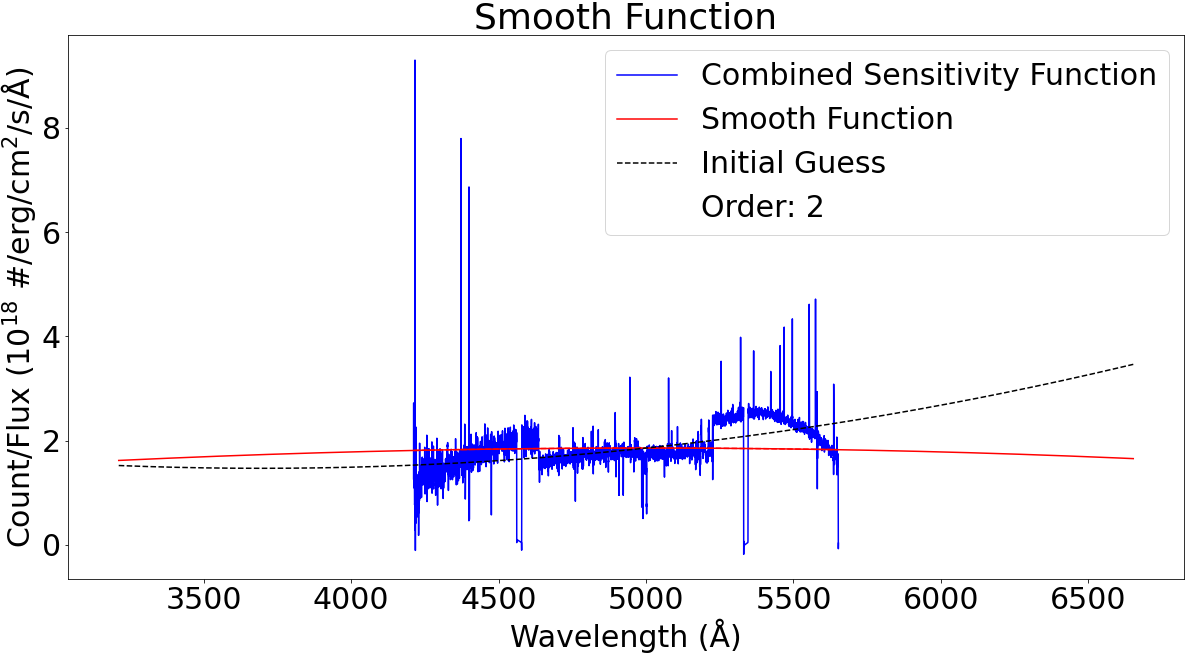}
   \caption{Plots showing diagnostic outputs from the \texttt{flux\_calibration} function for the PG2300 Dither$-$ observation. The left panel  shows both the individual sensitivity functions as well as their averaged combined sensitivity function. The right panel shows the same combined sensitivity function after having undergone polynomial regression. The resulting smoothed function (solid red line) is applied to the science images to remove influences from the jump discontinuities at the overlap region edges and the variance of the data.}
  \label{fig:sensfuncs}
\end{figure}

Before flux calibration is carried out, \texttt{CARRSSPipeline} makes some adjustments to the RSS observed spectra. First, the pipeline sets all pixels making up chip gaps in observed RSS spectra to NaN; this helps with averaging since \texttt{RSSMOSPipeline} outputs them as zero, which can affect the average process for removing chip gaps and systematics. \texttt{RSSMOSPipeline} offers a sky subtraction routine that is designed for science slits, since science is not typically done on alignment stars. 
Due to the bright stars dominating the sky background in boxes that are 4$''$ on a side, 
\texttt{RSSMOSPipeline} over-subtracts ``sky'' on our observed SDSS star spectra, but we require 
accurate 
sky subtraction 
since they are used to flux calibrate our science spectra. To address this, \texttt{CARRSSPipeline} measures the median values of the top and bottom 
rows 
of the star 
spectra, 
which are negative due to over-subtraction of sky. These 
negative 
values are then 
re-subtracted from 
the entire star spectra, bringing the edge values closer to zero and correcting for the missing flux.  

Once SDSS star directories are established, the routine makes a sensitivity function for each star by dividing the observed spectrum from SALT-RSS by the corresponding SDSS flux calibrated spectrum (see Figure~\ref{fig:obsvssdss}). The resulting functions in units of counts/flux are then averaged together to form a combined sensitivity function. This combined sensitivity function is then smoothed to reduce variance in the data, which is shown in Figure~\ref{fig:sensfuncs} for the PG2300 Dither$-$ observation. This is done using a regression routine from the Python package \texttt{AstroML}\footnote{https://github.com/astroML/astroML.git} \citep{astroML}, which outputs the second order polynomial coefficients for a smooth combined sensitivity function. However, instances of wavelength mismatch between the observed RSS star spectrum and the corresponding SDSS star spectrum can negatively affect solutions due to extrapolation of data at the mismatch wavelengths. This occurs at the far blue and red ends, where our observed RSS star spectra wavelengths do not completely overlap with the wavelength range that SDSS has for their spectra. To address this, we fit a polynomial to the sensitivity function using \texttt{SciPy}\footnote{https://github.com/scipy/scipy.git} \citep{2020SciPy-NMeth} routines and the \texttt{AstroML} coefficients as initial guesses. This would just output an equivalent solution to \texttt{AstroML}, but we include an additional constraint where the maximum value of this new function is set at a wavelength close to the maximum throughput of the setting. In the instance of wavelength mismatch, the initial guess and smooth function are no longer aligned, allowing the user to see how their custom polynomial compares to \texttt{AstroML} coefficients. In the instance of the PG2300 Dither$-$ observation for HETDEX J100041.45+021331.8, a maximum throughput constraint of 5000~\AA{} was needed. This approach helps the user fit a polynomial when mismatch occurs, and is based on the specific spectroscopic setting used for the data.
\startlongtable
  \begin{deluxetable}{c|lccccccccc} 
    \tabletypesize{\small}
    \tablecaption{Statistical Comparison Throughout Reduction Processes \label{tab:stats}}
    \tablehead{\colhead{ } & \colhead{ } & \colhead{ } & \colhead{ } & \colhead{ } & \colhead{$\Delta \lambda$ [\AA]} & \colhead{ }& \colhead{ } & \colhead{\tablenotemark{$^\ddagger$}$\Delta (log_{10}f)$ [dex]} & \colhead{\tablenotemark{$^\ddagger$}$\Delta (log_{10}f_{\lambda})$ [dex]}}
   \startdata
    \hline 
    \ Overlapping & Mean &&&& 0.064 &&& 0.002 & \hspace{-0.24cm}$-$0.131 \\
    \hline
    \ & Median &&&& \hspace{-0.24cm}$-$0.300 &&& 0.002 & \hspace{-0.24cm}$-$0.041  \\
    \hline
    & NMAD &&&& 1.038 &&& 0.044 & 0.128 \\
    \hline\hline
    Reprojected &   Mean && & & 0.150 &&& \hspace{-0.24cm}$-$0.017 & 0.098 \\
    \hline
    & Median && & &0.000 &&& \hspace{-0.24cm}$-$0.004 & \hspace{-0.24cm}$-$0.095 \\
    \hline
    & NMAD & && &0.148 &&& 0.059 & 0.121\\
    \enddata
    \tablenotetext{\ddagger}{\scriptsize Calculated using formula: $\Delta(log_{10}F) = log_{10}(F_{a}/F_{b})$, where $F_a, F_b$ are corresponding fluxes for spectra \textit{a},\textit{b}.}
    \tablecomments{\scriptsize Comparing measurements of emission lines and continua between overlapping settings as well as before and after reprojection for observed RSS science spectra. $f$ denotes fluxes of emission lines and $f_\lambda$ denotes median flux levels of continua. For comparisons between overlapping settings, negative values mean higher values for the bluer of the two settings. For comparisons before and after reprojection, negative values mean higher values for the combined reprojected spectrum. To convert to percent error, use \% Error = $(10^{\Delta(log_{10}F)}-1) \times 100$
  }
  \end{deluxetable}
After flux calibrating all settings for a particular target, we identified average offsets for wavelengths and fluxes of emission lines between overlapping settings. We additionally measure median continuum levels and compare those between settings (see Section \ref{subsec:cont} for details). Table \ref{tab:stats} shows these values, where we see some deviation for wavelengths, line fluxes, and continuum fluxes between overlapping settings post flux calibration. The median deviation is 0.002 dex for line fluxes and -0.041 dex for continuum levels. The larger scatter seen for the continuum levels is expected because they are more sensitive to noise and wavelength mismatch than emission lines at specific wavelengths. Overall, we consider the overlapping data to be in good agreement. 

\clearpage
\subsection{Full-Optical Spectroscopic Combination\label{subsec:speccombine}}

\begin{figure}[!h]
    \gridline{\fig{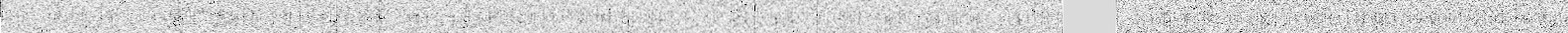}{\textwidth}{}}
    \vspace{-20pt}
    \gridline{\fig{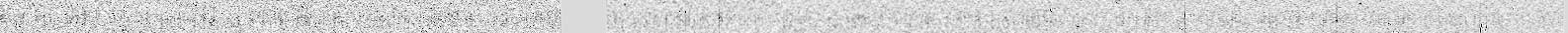}{\textwidth}{(a) PG3000}}
    \gridline{\fig{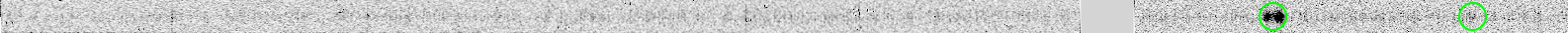}{\textwidth}{}}
    \vspace{-20pt}
    \gridline{\fig{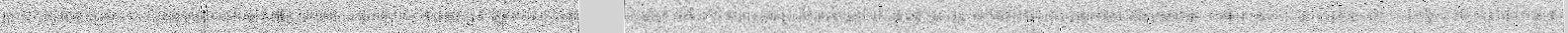}{\textwidth}{(b) PG2300}}
    \gridline{\fig{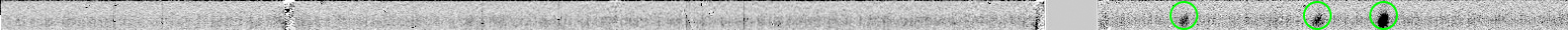}{\textwidth}{}}
    \vspace{-20pt}
    \gridline{\fig{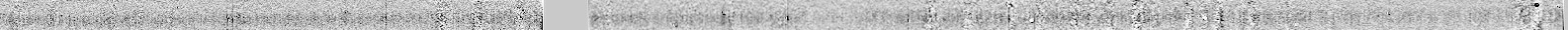}{\textwidth}{(c) PG0900}}
    \caption{Above are 2D spectroscopic images of HETDEX J100041.45+021331.8 post \texttt{RSSMOSPipeline} calibration, but prior to \texttt{CARRSSPipeline} in all Dither$-$ gratings. [O II] and [Ne III]$\lambda$3869 are visible in PG2300 Dither$-$, while [O III]$\lambda$4959,5007 and H$\beta$ are visible in PG0900 Dither$-$.}
    \label{fig:slit12postrss}
\end{figure}

Figure \ref{fig:slit12postrss} shows HETDEX J100041.45+021331.8 in all Dither$-$ settings after it has been run through the \texttt{RSSMOSPipeline}, but prior to the \texttt{CARRSSPipeline}. The [O II] doublet is visible in PG2300 and is resolved due to high SALT-RSS resolution. We can also see H$\beta$ and [O III]$\lambda$4959,5007 in PG0900. Following flux calibration of the data, these spectroscopic settings, along with their Dither+ counterparts, are combined on a single projected ``wavelength" axis yielding a fully reprojected, flux-calibrated 2D spectroscopic image of the target. The settings are reprojected onto an exponential function of wavelength, and we intentionally over-sample in the PG0900 settings to maintain (nearly) full resolution in the PG3000 and PG2300 settings.

\begin{equation}
  \lambda_{reproj} = A\;\mathrm{exp}\left({\frac{L}{k}+1}\right) 
  \label{eq:lam}
\end{equation}

  Equation \ref{eq:lam} shows the exponential function used in our reprojection, where L is a linear array of values that takes the place of a traditional wavelength array, k determines the rate at which resolution decreases toward longer wavelengths, and A is a scaling factor that causes the reprojected array to begin at the desired wavelength. The +1 serves as an additional parameter that causes the resolution per pixel ($\frac{\lambda}{d\lambda/dL}$) to drop by a factor of two from the bluest to reddest wavelength; adjusting that value would cause the resolution to change by a larger or smaller factor.
  
  \begin{figure}
    \plotone{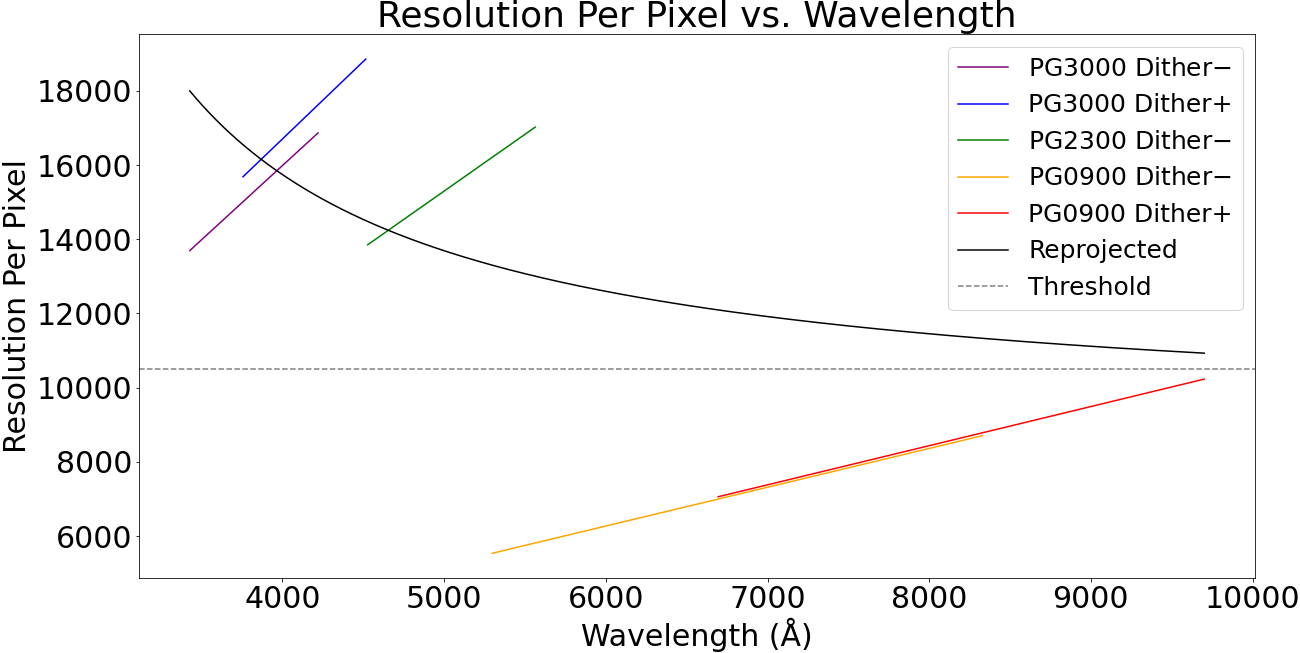}
    \caption{Resolution plot showing the relationship between resolution per pixel and wavelength. The colored lines represent each setting, with the color corresponding to the grating (violet/blue for PG3000, green for PG2300, orange/red for PG0900) and the sign representing the grating angle (Dither+ for the upper angle and Dither$-$ for the lower angle). The solid black line is the reprojected wavelength axis that the routine generates using an exponential fit. The dashed gray line is the minimum resolution threshold parameter used to favor higher resolution settings when they overlap with others.}
    \label{fig:resperpix}
\end{figure}

\begin{figure}[h]
    \plotone{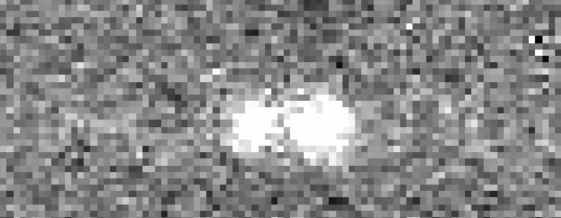}
    \caption{The [O II] doublet from HETDEX J100041.45+021331.8 after it has been flux calibrated and reprojected. The resolution of the [O II] doublet is well preserved.}
    \label{fig:oii2d}
\end{figure}

\begin{figure}
    \plotone{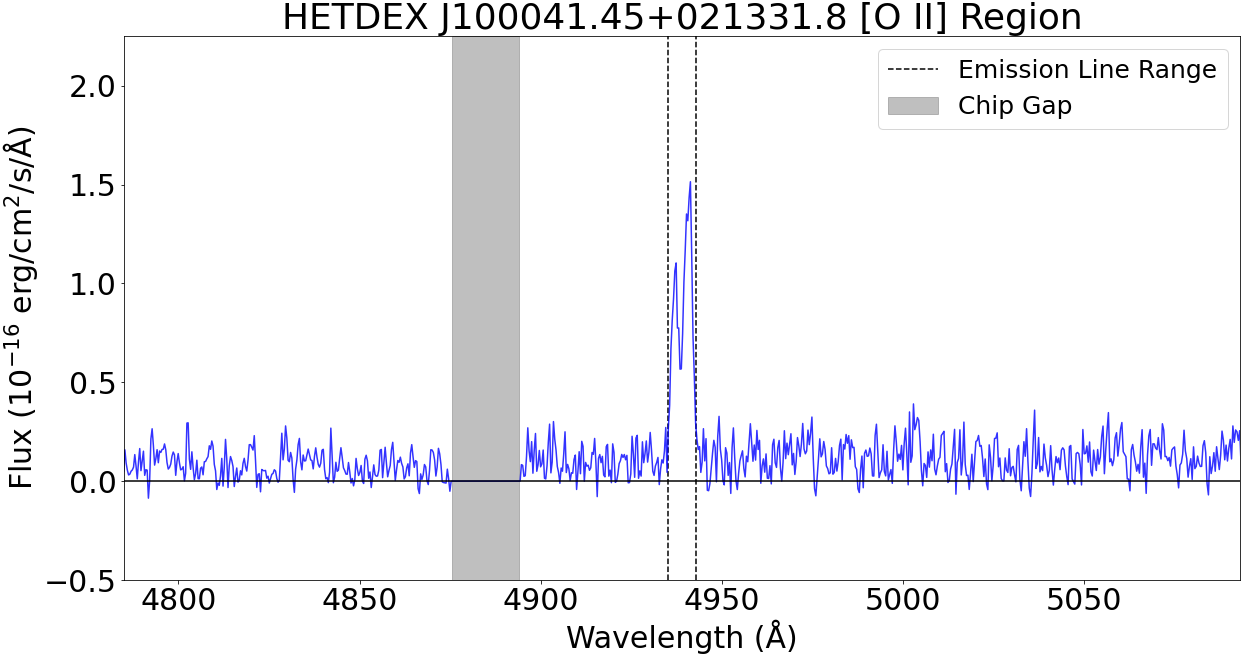}
    \caption{The [O II] doublet from HETDEX J100041.45+021331.8 in the reprojected spectrum at $\lambda$ = 4938.38\AA.}
    \label{fig:oii}
\end{figure}

  Figure \ref{fig:resperpix} illustrates the resolution vs. wavelength of the reprojected axis as well as the oversampling of the PG0900 settings. A potential limitation arises when data are present at overlapping wavelengths in both the PG2300 and PG0900 settings; since the resolution per pixel in the PG0900 settings is much lower than the PG3000 and PG2300 counterparts, features such as [O II] doublets lose resolution post reprojection. To prevent this, the \texttt{CARRSSPipeline} has a minimum resolution threshold that favors higher resolution settings. If there are two settings at a given wavelength above or below the threshold, they get averaged normally, otherwise if there is one below the threshold and one above the threshold, it will use the higher resolution setting. With this parameter, [O II] doublets retain their resolution as seen in Figures \ref{fig:oii2d} and \ref{fig:oii}. After reprojecting all settings for a particular target on a common exponential wavelength axis, we identified average offsets for wavelengths and fluxes of emission lines between individual spectra and the reprojected spectrum. We also measured median continuum levels and compared those between individual spectra and the reprojected spectrum (see Section \ref{subsec:cont} for details). Table \ref{tab:stats} shows these values, with good agreement of -0.004 dex for line fluxes and decent agreement of -0.095 dex for continua.

\subsection{1D Extraction\label{subsec:1D}}
\begin{figure}
    \plotone{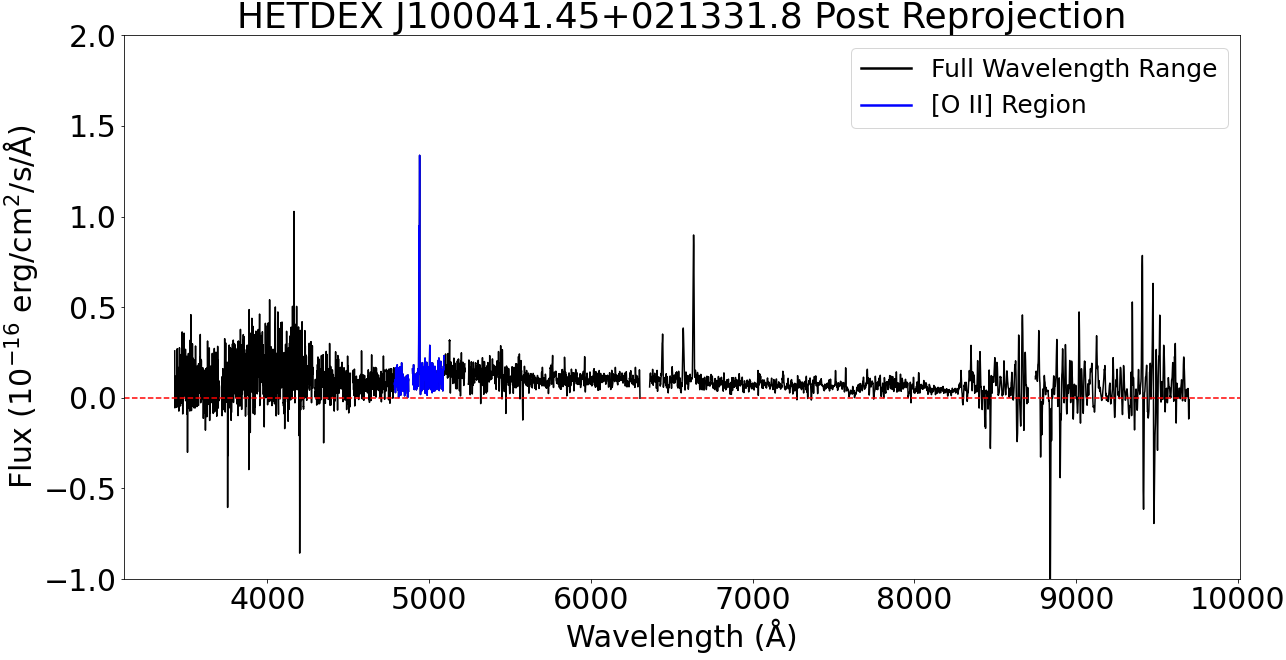}
    \caption{This plot illustrates the 1D flux calibrated, reprojected spectrum of HETDEX J100041.45+021331.8 after Gaussian smoothing with a sigma = 1.5 pixel kernel. The bluest setting is noisy but the continuum level still matches HETDEX relatively well. The [O II] region shown in Figure \ref{fig:oii} is highlighted in blue.}
    \label{fig:1dsig}
\end{figure}

After the \texttt{CARRSSPipeline} outputs a fully combined flux-calibrated 2D spectrum, the target signal is extracted and saved as a 1D spectrum. We utilize the \texttt{finalExtration} function on \texttt{RSSMOSPipeline} to obtain trace centers and sigmas for each column, with a moving window of several hundred pixels. The trace centers and sigmas generate Gaussian fits across the spatial axis of the 2D RSS science spectra. The \texttt{finalExtration} function also offers a linear running profile made using the trace centers and sigmas, but we instead implemented a custom trace fitter in the \texttt{CARRSSPipeline} that makes a running profile using a polynomial fit of order 4. Finally, we multiply the 2D RSS science spectra with this profile and sum the weighted rows to extract the 1D spectrum. Figure \ref{fig:1dsig} shows the \texttt{CARRSSPipeline} outputted 1D spectrum for HETDEX J100041.45+021331.8 spanning the full wavelength coverage.  The pipeline allows the user to select which settings they want to include given the context of their specific program.

\clearpage
\startlongtable
  \begin{deluxetable}{lccccc} 
    \tabletypesize{\small}
    \tablecaption{Flux Statistics Between RSS and HETDEX Spectra \label{tab:stat}}
    \tablehead{\colhead{ } &\colhead{ } & \colhead{ } & \colhead{ } &
      \colhead{\tablenotemark{$^\ddagger$}$\Delta (log_{10}f)$ [dex]} & \colhead{\tablenotemark{$^\ddagger$}$\Delta (log_{10}f_{\lambda})$ [dex]} }
   \startdata
    Mean &&&& \hspace{-0.24cm}$-$0.292 & \hspace{-0.24cm}$-$0.367  \\
    \hline
    Median &&&& \hspace{-0.24cm}$-$0.286 & \hspace{-0.24cm}$-$0.358   \\
    \hline
    NMAD &&&& 0.064 & 0.141 
    \enddata
    \tablenotetext{\ddagger}{\scriptsize $\Delta(log_{10}F) = log_{10}(F_{\mathrm{rss}}/F_{\mathrm{hetdex}})$}
    \tablecomments{\scriptsize Comparing statistical offsets of emission lines and continua between our observed RSS science spectra and HETDEX. $f$ denotes fluxes of emission lines and $f_\lambda$ denotes median flux levels of continua. Positive values denote overestimation and negative values denote underestimation compared to HETDEX.
  }
  \end{deluxetable}

Table \ref{tab:stat} shows how the flux calibration of RSS observed science spectra compares to that of HETDEX for emission lines fluxes ($f$) and median flux values of the fitted continuum ($f_\lambda$) across the wavelength range that overlaps with HETDEX coverage (see Section \ref{subsec:cont} for details on continua measurements). These measurements indicate a multiplicative bias for our program in reference to HETDEX
of -0.286 dex. 
We attribute this factor of $\sim2$ flux loss to a combination of astrometric errors, imperfect mask alignment, and atmospheric seeing.  
All flux data reported in subsequent tables and flux-calibrated spectra shown in figures in this paper have already been corrected for this bias, which we implement in the \texttt{CARRSSPipeline}. The pipeline defaults this value to one until the user can identify their own offset dependant on their specific program. 

\subsection{Continuum Comparison\label{subsec:cont}}

\begin{figure}[h]
    \plotone{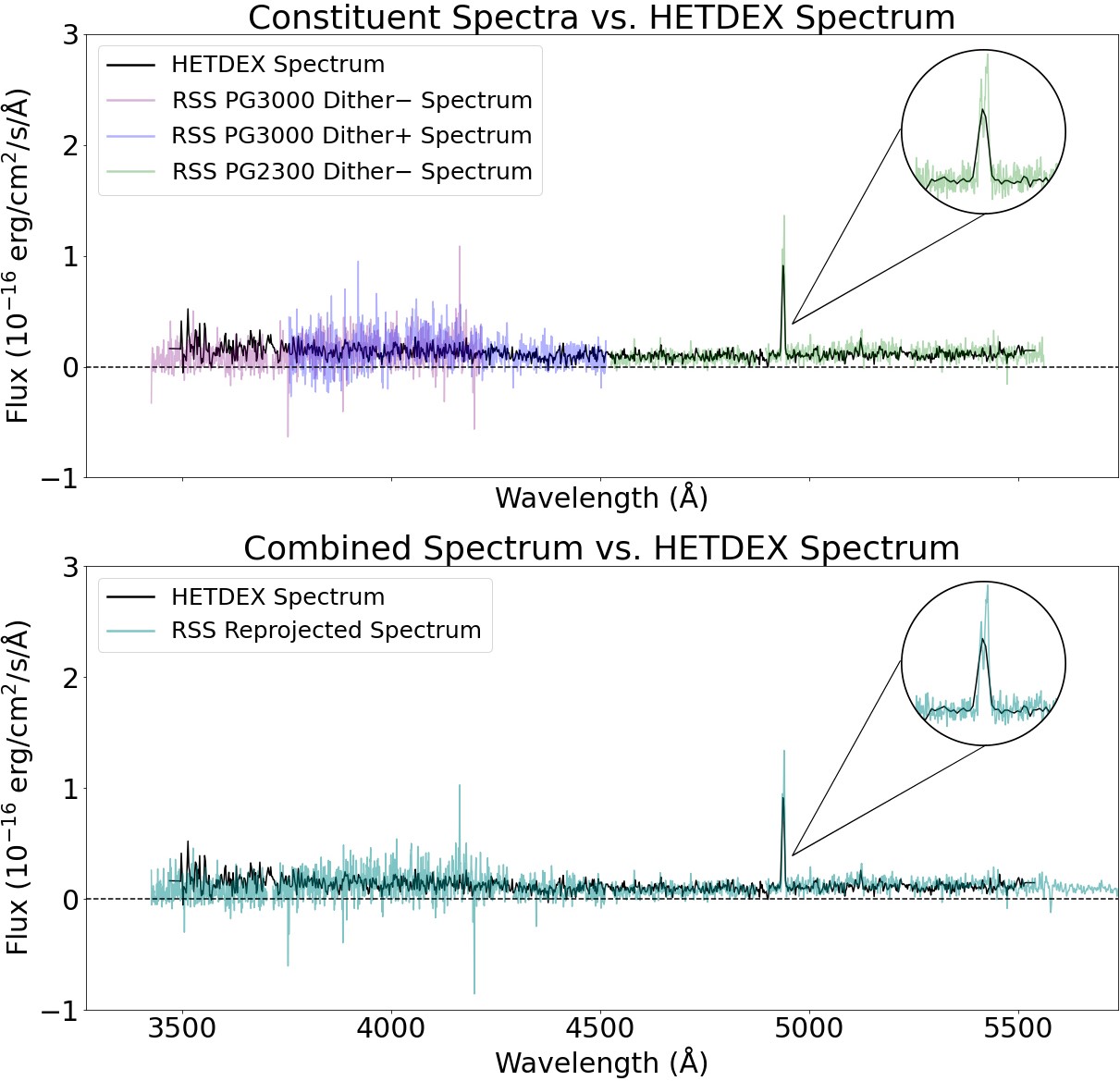}
    \caption{Plots outputted from the spectral comparison function showing all the extracted signals of spectra from each setting (upper plot) and the combined spectrum (lower plot) compared to the HETDEX spectrum for J100041.45+021331.8 after Gaussian smoothing with a sigma = 1.5 pixel kernel. Note that the [O II] doublets overlap but is unresolved in the HETDEX spectrum (R $\sim$ 800).}
    \label{fig:cont}
\end{figure}

\startlongtable
  \begin{deluxetable}{cccccc} 
    \tabletypesize{\normalsize}
    \tablewidth{\textwidth}
    \tablecaption{Flux Measurements for Continua \label{tab:cont}}
    \tablehead{
      \colhead{Target Name} & \colhead{} & \colhead{}&\colhead{}&\colhead{\tablenotemark{$^\dagger$}$f_{\lambda,\mathrm{rss}}$}  & \colhead{\tablenotemark{$^\ddagger$}$\Delta(log_{10}f_\lambda)$ [dex]} 
    }
   \startdata
    \begin{filecontents*}{continuum_table_v2.csv}
,,,,,
HETDEX J100057.95+021524.4, , , ,0.025,$-$0.512
HETDEX J100048.38+021454.2, , , ,0.361,\hspace{0.3cm}0.023
HETDEX J100032.77+021357.3, , , ,0.229,$-$0.128
HETDEX J100041.45+021331.8, , , ,0.133,\hspace{0.3cm}0.060
\end{filecontents*}
\csvreader[head to column names,%
        late after line=\\,%
        ]{continuum_table_v2.csv}{}
    {\csvcoli & \csvcolii & \csvcoliii & \csvcoliv & \csvcolv & \csvcolvi} 
    \enddata
    \tablenotetext{\dagger}{\scriptsize In units of 10$^{-16}$ ergs~cm$^{-2}$~s$^{-1}$}
    \tablenotetext{\ddagger}{\scriptsize Calculated using formula: $\Delta(log_{10}f_{\lambda})$ = $log_{10}(f_{\lambda, \mathrm{rss}}/f_{\lambda, \mathrm{hetdex}})$}
    \tablecomments{\scriptsize Fluxes here denote median values of the fitted continuum across the wavelength range that overlaps with HETDEX coverage. Positive (negative) values in the final column denote overestimation (underestimation).
  }
  \end{deluxetable}

If the observed source has an available spectrum in the literature to compare with, the user can take advantage of the spectral comparison function \texttt{continuum\_comparison} offered by the \texttt{CARRSSPipeline}. In addition to comparing continua, it also plots the extracted signal for each setting and for the combined setting, all post flux calibration. This allows the user to visually compare the continua of reduced spectra against literature spectra across the same wavelength range. Figure \ref{fig:cont} shows these outputs for HETDEX J100041.45+021331.8, where we can further confirm the wavelength calibration accuracy by seeing the [O II] doublets overlapped. These continuum comparisons have been carried out for all targets in this paper, and those results are shown in Table \ref{tab:cont}. The larger disagreement for HETDEX J100057.95+021524.4 comes from the bluest observation, where the flux calibration was significantly affected by poor data quality at low wavelengths. Overall, the median continuum values between our RSS observed spectra and HETDEX agree to an average of $-$0.14 dex.

\subsection{Line Visual Inspection \label{subsec: line}}
All targets in this paper have been observed by HETDEX, allowing us to compare line fluxes and wavelengths for [O II] doublets and other available lines. The \texttt{CARRSSPipeline} offers a function called \texttt{line\_inspection} that takes in the emission line wavelength range and a configuration file with known emission line rest wavelengths. Using these, the centroid of the emission line is calculated using the centroid function offered by another Python package called \texttt{Specutils}\footnote{https://specutils.readthedocs.io/en/stable/} \citep{nicholas_earl_2024_11099077}, and used to measure the observed wavelength and redshift of our RSS observed science spectra. Before line flux measurements are carried out, the spectra are first subject to continuum subtraction, which takes advantage of a continuum fitter from \texttt{Specutils}. The \texttt{line\_inspection} function also uses the \texttt{line\_flux} function from \texttt{Specutils}, which takes the integrated flux of the emission line over the emission line wavelength range and outputs both the line flux and an error that is based on an empirical method of fitting the spectrum to a polynomial and measuring the root mean squared of the signal at each point. A limitation of this approach arises due to systematics that causes spikes by poorly subtracted sky lines, but we manage this limitation by using a moving window of 100 pixels to measure the RMS at each pixel. Figure \ref{fig:oii} shows a 1D extraction of the [O II] doublet from HETDEX J100041.45+021331.8 made from this module. Table \ref{tab:emlines} shows the emission line wavelength and line flux measurements for each target. Wavelength calibrations for all targets in this paper are good to within $\sim\pm$300 km s$^{-1}$ ($\Delta z = 0.001$), with most being good to $\sim\pm$30 km s$^{-1}$ ($\Delta z = 0.0001$). 

\centerwidetable
\startlongtable
  \begin{deluxetable*}{cccccccccc} 
    \tabletypesize{\small}
    \tablewidth{0pt}
    \tablecaption{Wavelength and Flux Measurements for Emission Lines \label{tab:emlines}}
    \tablehead{\colhead{Target Name} & \colhead{Line}  & \colhead{Confidence} & \colhead{$\lambda_{\mathrm{rss}}$ [\AA]}&
    \colhead{\tablenotemark{*}$\sigma_\lambda$ [\AA]}&
    \colhead{$z$} &  \colhead{\hspace{0.3cm}$\Delta z$} & \colhead{\tablenotemark{$^\dagger$}$f_{\mathrm{rss}}$}&
    \colhead{\tablenotemark{$^\dagger$}$\sigma_f$}&
    \colhead{\tablenotemark{$^\ddagger$}$\Delta(log_{10}f)$} [dex]}
    \startdata
    \begin{filecontents*}{Line_table_v3.csv}
,,,,,,,,,
HETDEX J100057.95+021524.4,[O II],2,5030.3,1.7,1.34954,\hspace{0.3cm}0.00062,3.172,0.203,$-$0.118
HETDEX J100048.38+021454.2,[O II],3,4074.0,0.8,1.09297,\hspace{0.3cm}0.00017,7.276,0.195,\hspace{0.3cm}0.049
HETDEX J100048.38+021454.2,H$\beta$,3,5313.1,0.4,1.09293,\hspace{0.3cm}0.00013,1.997,0.056,$-$0.063
HETDEX J100048.38+021454.2,[O III]$\lambda$5007,2,5471.3,0.5,1.09277,$-$0.00003,1.436,0.064,\hspace{0.3cm}0.011
HETDEX J100032.77+021357.3,[O II],3,4184.9,1.3,1.12273,\hspace{0.3cm}0.00064,9.771,0.138,$-$0.061
HETDEX J100041.45+021331.8,[O II],3,4938.9,1.6,1.32500,\hspace{0.3cm}0.00086,5.882,0.153,\hspace{0.3cm}0.014
\end{filecontents*}
\csvreader[head to column names,%
        late after line=\\,%
        ]{Line_table_v3.csv}{}
    {\csvcoli & \csvcolii & \csvcoliii & \csvcoliv & \csvcolv & \csvcolvi & \csvcolvii & \csvcolviii & \csvcolix & \csvcolx} 
    \enddata
    \tablenotetext{*}{\scriptsize Median offsets outputted by \texttt{RSSMOSPipeline} by comparing wavelength solution to a dictionary of sky lines.}
    \tablenotetext{\dagger}{\scriptsize In units of 10$^{-16}$ ergs~cm$^{-2}$~s$^{-1}$}
    \tablenotetext{\ddagger}{\scriptsize Calculated using formula: $\Delta(log_{10}f)$ = $log_{10}(f_{\mathrm{rss}}/f_{\mathrm{hetdex}})$}
    \tablecomments{\scriptsize 
    Confidence denotes how visible the emission line is in both the 2D and 1D spectra. 3 is completely visible and distinct from surrounding noise; many 3s are visible before reduction even begins. 2 is clearly visible, slightly noisier and may appear less Gaussian. These may be hard to see prior to reduction. 1 denotes that something is there, but difficult to see even after reduction.
    \\
    Positive values denote overestimation and negative values denote underestimation compared to HETDEX.
}
  \end{deluxetable*}

\section{Conclusions \label{sec:conc}}
With the goal of expanding the scientific capabilities of SALT-RSS, our \texttt{CARRSSPipeline} is designed to minimize a physical limitation that arises from the variable primary mirror illumination seen by the tracker. By using SDSS stars with 
calibrated spectra as alignment stars, our program allows astronomers to flux calibrate science spectra. The pipeline follows with reprojection and 1D extraction routines that output fully reprojected science spectra ready for line flux and continuum analysis that \texttt{CARRSSPipeline} also offers.

The pipeline is sensitive to RSS observed star spectra quality; our COSMOS-mask-B observes alignment stars in square boxes as opposed to slits. To minimize the risk of vertical 
and horizontal 
slit losses, we suggest that these calibration stars be put in
roughly 
10\arcsec{} high x 4\arcsec{} wide slits 
when possible, 
otherwise we suggest making those slits the same size as the science slits to ensure better overall sky subtraction and flux calibration. Slit losses appear to be affecting our science observations, due to a combination of PSF width, astrometric errors in mask design, and imperfect mask alignment. 
The 
median line flux error of $-$0.286 dex 
seen in Table \ref{tab:stat}
represents 
a
multiplicative bias for our program, which we 
then 
correct for during flux calibration.
All flux data reported in this paper have already been corrected for this bias, as well as all figures that show flux-calibrated spectra.  After correcting for this bias, the \texttt{CARRSSPipeline} is capable of flux calibration and reprojection with a 0.064 dex accuracy,
as reflected by the normalized median absolute deviation. 

After reducing all 
observational 
data, 
analyses of emission lines will shed light on various galaxy characteristics for these [O II]-emitters. A follow-up paper detailing flux analyses and science will follow the reduction of all masks at our disposal, giving us an improved view of [O II]-emitting galaxies.

\begin{acknowledgements}
\section{Acknowledgements}
We would like to thank the Department of Physics \& Astronomy at Rutgers University, SALT staff, and collaborators from ODIN for their support. We  acknowledge support from NSF grants AST-2206222 and AST-2206705, as well as NASA Astrophysics Data Analysis Program grant 80NSSC22K0487.

This material is based upon work supported by the National Science Foundation Graduate Research Fellowship Program under Grant No. DGE-2233066 to NF. NF would also like to thank the LSST-DA Data Science Fellowship Program, which is funded by LSST Discovery Alliance, NSF Cybertraining Grant 1829740, the Brinson Foundation, and the Moore Foundation; her participation in this program has benefited this work. 
\end{acknowledgements}

%

\vspace{5mm}
\facilities{SALT-RSS}


\software{\texttt{Astropy} \citep{astropy:2013,astropy:2018,astropy:2022}, \texttt{AstroML} \citep{astroML}, Principal Investigator Proposal Tool (PIPT), \texttt{PySALT} \citep{10.1117/12.857000}, \texttt{RSSMOSPipeline} \citep{Hilton_2018}, \texttt{SciPy} \citep{2020SciPy-NMeth}, \texttt{Specutils} \citep{nicholas_earl_2024_11099077}
          }




\bibliography{PASPsample631}{}
\bibliographystyle{aasjournal}



\end{document}